\documentclass[11pt,floatfix,noshowpacs]{revtex4}
\usepackage[dvips]{graphicx,color}

\usepackage[caption=false]{subfig}

\usepackage[latin1]{inputenc}
\usepackage{amsmath,amssymb,amsfonts,amsthm,latexsym}
\usepackage[english]{babel}
\usepackage{float}
\usepackage{subfig}
\usepackage{slashed}

\begin{document}

\author{J.P. Carlomagno$^{a,b}$, D. G\'omez Dumm$^{a,b}$ and N.N.\ Scoccola$^{b,c,d}$}

\address{$^{a}$ IFLP, CONICET $-$ Dpto.\ de F\'{\i}sica, Universidad
Nacional de La Plata, C.C. 67, 1900 La Plata, Argentina,\\
$^{b}$ CONICET, Rivadavia 1917, 1033 Buenos Aires, Argentina \\
$^{c}$ Physics Department, Comisi\'on Nacional de Energ\'{\i}a Atómica,
Av.Libertador 8250, 1429 Buenos Aires, Argentina \\
$^{d}$ Universidad Favaloro, Sol{\'\i}s 453, 1078 Buenos Aires, Argentina}

\title{\sc\Large{Deconfinement and chiral restoration in nonlocal SU(3) chiral quark models}}

\begin{abstract}
We study the features of nonlocal SU(3) chiral quark models with wave
function renormalization. Model parameters are determined from meson
phenomenology, considering different nonlocal form factor shapes. In this
context we analyze the characteristics of the deconfinement and chiral
restoration transitions at finite temperature, introducing the couplings of
fermions to the Polyakov loop. We analyze the results obtained for various
thermodynamical quantities considering different Polyakov loop potentials
and nonlocal form factors, in comparison with data obtained from lattice
QCD calculations.
\end{abstract}

\maketitle


\section{\sc Introduction}

The detailed understanding of the behavior of strongly interacting matter
under extreme conditions of temperature and/or density has become an issue
of great interest in recent years. It is widely believed that as the
temperature and/or density increase, one finds a transition from a hadronic
phase, in which chiral symmetry is broken and quarks are confined, to a
partonic phase in which chiral symmetry is restored and/or quarks are
deconfined. From the theoretical point of view, one way to address this
problem is through lattice QCD calculations~\cite{All03,Fod04,Kar03}, which
have been significantly improved in the last years. However, this {\em ab
initio} approach is not yet able to provide a full understanding of the QCD
phase diagram and the related hadron properties, owing to the well-known
difficulties of dealing with small current quark masses and finite chemical
potentials. Thus, it is worth developing effective models that show
consistency with lattice results, and can be extrapolated into regions not
accessible by lattice calculation techniques. Here we will concentrate on
one particular class of effective theories, namely the so-called nonlocal
Polyakov$-$Nambu$-$Jona-Lasinio (nlPNJL)
models~\cite{Blaschke:2007np,Contrera:2007wu,Contrera:2009hk,Hell:2008cc,Hell:2009by},
in which quarks move in a background color field and interact through
covariant nonlocal chirally symmetric four-point couplings. These
approaches, which can be considered as an improvement over the (local) PNJL
model~\cite{Meisinger:1995ih,Fukushima:2003fw,Megias:2004hj,Ratti:2005jh,
Roessner:2006xn,Mukherjee:2006hq,Sasaki:2006ww}, offer a common framework to
study both the chiral restoration and deconfinement transitions. In fact,
the nonlocal character of the interactions arises naturally in the context
of several successful approaches to low-energy quark
dynamics~\cite{Schafer:1996wv,RW94}, and leads to a momentum dependence in
the quark propagator that can be made consistent~\cite{Noguera:2008} with
lattice results~\cite{bowman,Parappilly:2005ei,Furui:2006ks}. Moreover, it
has been found that, under certain conditions, it is possible to derive the
main features of nlPNJL models starting directly from
QCD~\cite{Kondo:2010ts}.

Some previous works have addressed the study of nlPNJL models for the case
of two dynamical quarks, showing that the presence of nonlocal form factors
in the current-current quark interactions leads to a momentum-dependent mass
and wave function renormalization (WFR) in the quark
propagator~\cite{Contrera:2010kz,Pagura:2011rt,Pagura:2012ku}. As stated, it
is possible to choose the model parameters and form factors so as to fit
these momentum dependences to those obtained in lattice
QCD~\cite{Noguera:2008}. The aim of this work is to extend those works to
three flavors, including flavor mixing through a nonlocal 't Hooft-like
six-fermion interaction. The case of a three-flavor nlPNJL model with simple
Gaussian form factors and no WFR in the quark propagator has been previously
addressed in Refs.~\cite{Contrera:2007wu,Contrera:2009hk}, where the
phenomenology of light scalar and pseudoscalar mesons is analyzed. In
addition, the introduction of a Gaussian form factor to account for the WFR
in the three-flavor case has been considered in Ref.~\cite{Hell:2011ic}. For
comparison, we analyze here both the case of a model in which the form
factors are Gaussian functions (which ensure a fast ultraviolet convergence
of loop integrals), and a model in which these are given by the mentioned
lattice QCD-inspired functions [see Eqs.~(\ref{ff2},\ref{faux2}) below] of
the momentum. In this framework we determine several properties of light
mesons (masses, mixing angles, decay constants), analyzing the compatibility
with the corresponding phenomenological values. Then we study the
deconfinement and chiral restoration phase transitions that occur at finite
temperature, and we determine the corresponding critical temperatures. Our
analyses are carried out at the mean field level, considering the
above-mentioned form factor shapes and various parameter sets and functional
forms for the Polyakov potential. We also analyze the behavior of
thermodynamical quantities such as the interaction energy and the entropy
and energy densities. The results are discussed in comparison with data
obtained from lattice QCD calculations.

This article is organized as follows. In Sec.\ II we present the general
formalism, including analytical results for the scalar and pseudoscalar
meson properties. We discuss the model parameterization and compare our
predictions with phenomenological expectations. In Sec.\ III we extend our
analysis to nonzero temperature. The Polyakov loop potential is introduced
and the deconfinement and chiral restoration phase transitions are analyzed.
In Sec.\ IV we summarize our results and conclusions. The Appendix includes
some of our analytical expressions.

\section{\sc Nonlocal SU(3) chiral quark model - Zero temperature}

We start by considering the Euclidean effective action
\begin{eqnarray}
\label{se}
S_E &=& \int d^4x \ \left\{ \overline{\psi}(x)(-\imath \slashed D +
\hat m)\psi(x)-\frac{G}{2}\left[
j_a^S(x)j_a^S(x)+j_a^P(x)j_a^P(x)+j^r(x)j^r(x)\right] \right.
\nonumber\\
    &&\left. -\frac{H}{4} A_{abc}\left[
j_a^S(x)j_b^S(x)j_c^S(x)-3j_a^S(x)j_b^P(x)j_c^P(x)\right] \ + \ {\cal U}
\,[{\mathcal A}(x)]
\right\} \ ,
\end{eqnarray}
where $\psi(x)$ is the $N_f=3$ fermion triplet $\psi = (u\ d\ s)^T$, and
$\hat m={\rm diag}(m_u,m_d,m_s)$ is the current quark mass matrix. We will
work in the isospin symmetry limit, assuming $m_u=m_d$. The fermion
currents are given by
\begin{eqnarray}
j_a^s(x) &=& \int d^4z\; g(z)\,
\overline{\psi}\left(x+\frac{z}{2}\right)\lambda_a
\psi\left(x-\frac{z}{2}\right)
\ , \nonumber\\
j_a^p(x) &=& \int d^4z\; g(z)\,
\overline{\psi}\left(x+\frac{z}{2}\right)\imath \lambda_a \gamma_5
\psi\left(x-\frac{z}{2}\right)
\ , \nonumber\\
j^r(x)   &=& \int d^4z\; f(z)\, \overline{\psi}\left(x+\frac{z}{2}\right)
\frac{\imath \overleftrightarrow{\slashed \partial}}{2\kappa}
\psi\left(x-\frac{z}{2}\right)\ ,
\end{eqnarray}
where $f(z)$ and $g(z)$ are covariant form factors responsible for the
nonlocal character of the interactions, and $\lambda_a$, $a=0,...,8$, are
the standard eight Gell-Mann matrices, plus
$\lambda_0=\sqrt{2/3}\,\mathbf{1}_{3\times 3}$. The relative weight of the
interaction driven by $j^r(x)$, which is responsible for the quark wave
function renormalization, is controlled by the parameter $\kappa$. The model
includes flavor mixing through a 't Hooft-like term, in which the SU(3)
symmetric constants $A_{abc}$ are defined by
\begin{equation}
\label{aabc}
A_{abc}=\frac{1}{3!}\epsilon_{ijk}\epsilon_{mnl}(\lambda_a)_{im}
(\lambda_b)_{jn}(\lambda_c)_{kl} \ .
\end{equation}
The interaction between fermions and color gauge fields $G_\mu^a$ takes
place through the covariant derivative in the fermion kinetic term,
$D_\mu\equiv
\partial_\mu - \imath {\mathcal A}_\mu$, where ${\mathcal A}_\mu = g\,
G_\mu^a \lambda^a/2$. Finally, the action includes an effective potential
${\cal U}$ that accounts for gauge field self-interactions. At the mean
field level we will assume that fermions move on a uniform background gauge
field, which for zero temperature decouples from matter (finite temperature
effects will be discussed in the next sections).

To work with mesonic degrees of freedom we proceed to perform a standard
bosonization of the fermionic theory, introducing scalar fields
$\sigma_a(x)$, $\zeta(x)$ and pseudoscalar fields $\pi_a(x)$, together
with auxiliary fields $S_a(x)$, $P_a(x)$ and $R(x)$, with $a=0,...,8$.
After integrating out the fermion fields we obtain a partition function
\begin{eqnarray}
\label{zbos}
\mathcal{Z} &=& \int \mathcal{D}\sigma_a\mathcal{D}\pi_a\mathcal{D}\zeta\
A(\sigma_a,\pi_a,\zeta)
\nonumber\\
&\times& \int \mathcal{D}S_a \mathcal{D}P_a \mathcal{D}R \ \exp
\int d^4x \bigg[ \sigma_a S_a + \pi_a P_a +\zeta
R+\frac{G}{2}(S_aS_a+P_aP_a+R^2) +
\nonumber\\
& + & \frac{H}{4}A_{abc} (S_a S_b S_c - 3 S_a P_b P_c) \bigg] \ ,
\end{eqnarray}
where the operator $A(p,p^\prime)$ (in momentum space) is given by
\begin{eqnarray}
\label{A}
A(p,p^\prime)&=&(2\pi)^4 \delta^{(4)}(p-p^\prime)(-\slashed p+m_c) +
g\left(\frac{p+p^\prime}{2}\right)[\sigma_a(p^\prime-p)+\imath \gamma_5
\pi_a(p^\prime-p)]\lambda_a +
\nonumber\\
&+&\frac{1}{2\kappa}f\left(\frac{p+p^\prime}{2}\right)(\slashed p+\slashed p^\prime)
\zeta(p^\prime-p)\ .
\end{eqnarray}
Now we follow the stationary phase approximation, replacing the path
integrals over the auxiliary fields by the corresponding argument
evaluated at the minimizing values $\tilde{S_a}$, $\tilde{P_a}$, and
$\tilde{R}$. The procedure is similar to that carried out in
Ref.~\cite{Scarpettini:2003fj}, where more details can be found.

\subsection{\sc Mean Field Approximation}
We consider the mean field approximation (MFA), in which the meson
fields are expanded around their vacuum expectation values. One thus has
\begin{eqnarray}
\label{mfaf}
\sigma_a(x) &=& \bar \sigma_a+\delta\sigma_a(x) \ , \nonumber\\
\pi_a(x) &=& \delta\pi_a(x) \ , \nonumber\\
\zeta(x) &=& \bar\zeta+\delta\zeta(x) \ ,
\end{eqnarray}
where we have assumed that pseudoscalar mean field values vanish, owing to
parity conservation. Moreover, for the scalar fields only
$\bar\sigma_{0,8}$ and $\bar\zeta$ can be different from zero due to
charge and isospin symmetries. Thus, the Euclidean action reduces to
\begin{equation}
\label{semfa}
\frac{S_E^{\mbox{\tiny MFA}}}{V^{(4)}} =  -2\; {\rm Tr}\; \int \frac{d^4p}{(2\pi)^4}
\log\left[\frac{M^2(p)+p^2}{Z^2(p)}\right]
- \bar\sigma_a\bar S_a-\bar\zeta\bar R-
\frac{G}{2}(\bar S_a\bar S_a+\bar R^2)-
\frac{H}{4}A_{abc}\bar S_a\bar S_b\bar S_c \ ,
\end{equation}
where $\bar S_a$, $\bar P_a$, and $\bar R$ stand for the values of
$\tilde{S_a}$, $\tilde{P_a}$ and $\tilde{R}$ within the MFA.

For the neutral fields ($a=0,3,8$) it is convenient to change to a flavour
basis, $\phi_a \to \phi_i$, where $i=u,d,s$, or equivalently $i=1,2,3$.
In this basis, by minimizing the mean field action in Eq.~(\ref{semfa}) we
obtain the gap equations given in Ref.~\cite{Scarpettini:2003fj},
\begin{eqnarray}
\bar \sigma_u + G\bar S_u+\frac{H}{2}\bar S_d \bar S_s &=& 0 \ , \nonumber\\
\bar \sigma_d + G\bar S_d+\frac{H}{2}\bar S_s \bar S_u &=& 0 \ , \nonumber\\
\bar \sigma_s + G\bar S_s+\frac{H}{2}\bar S_u \bar S_d &=& 0 \ ,
\label{gaps}
\end{eqnarray}
plus an extra equation arising from the $j^r(x)$ current-current
interaction,
\begin{equation}
\bar\zeta +G\bar R  =  0 \ ,
\end{equation}
where the mean field values $\bar S_i$ and $\bar R$ are given by
\begin{eqnarray}
\bar S_i &=& -8N_c \int \frac{d^4p}{(2\pi)^4}\ g(p)\;
\frac{Z(p)\, M_i(p)}{p^2 + M^2_i(p)} \ , \ i=u,d,s, \nonumber\\
\bar R &=& \frac{4 N_c}{\kappa} \int \frac{d^4p}{(2\pi)^4}\ p^2\, f(p)
\; \sum_{i=1}^3 \frac{Z(p)}{p^2 + M^2_i(p)} \ .
\label{sr}
\end{eqnarray}
The functions $M_i(p)$ and $Z(p)$ correspond to momentum-dependent effective
masses and WFR of the quark propagators. In terms of the model parameters
and form factors, these are given by
\begin{eqnarray}
M_i(p) &=& Z(p)\, \left[ m_i\, +\, \bar \sigma_i\, g(p)\right] \ ,
\nonumber\\
Z(p) &=& \left[ 1\,-\,\frac{\bar \zeta}{\kappa}\,f(p)\right]^{-1}\ .
\label{mz}
\end{eqnarray}
Thus, for a given set of model parameters and form factors, from
Eqs.~(\ref{gaps}-\ref{mz}) one can numerically obtain the mean field
values $\bar \sigma_{u,s}$ and $\bar\zeta$.

The chiral condensates $\langle \bar q q\rangle$, order parameters of the
chiral restoration transition, can be obtained by varying the MFA partition
function with respect to the current quark masses. These quantities are, in
general, divergent and can be regularized by subtracting the free quark
contributions. One has
\begin{equation}
\langle \bar q q\rangle \ = \ -4 N_c \int \frac{d^4p}{(2\pi)^4}
\left[\frac{Z(p)M_q(p)}{p^2 + M^2_q(p)}\; -\; \frac{m_q}{p^2+m_q^2}\right]
, \ q = u,d,s.
\end{equation}

\subsection{\sc Quadratic Fluctuations - Meson masses and weak decay
constants} \label{qf}

In order to analyze the properties of meson fields it is necessary to go
beyond the MFA, considering quadratic fluctuations in the Euclidean
action:
\begin{eqnarray}
\label{spiketa}
S_E^{\rm quad} &=& \dfrac{1}{2} \int \frac{d^4 p}{(2\pi)^4} \sum_{M}\  r_M\
G_M(p^2)\  \phi_M(p)\, \bar\phi_M(-p) \ ,
\end{eqnarray}
where meson fluctuations $\delta\sigma_a$, $\delta\pi_a$  have been
translated to a charge basis $\phi_M$, $M$ being the scalar and
pseudoscalar mesons in the lowest mass nonets ($\sigma,\pi^0$, etc.), plus
the $\zeta$ field. The coefficient $r_M$ is 1 for charge eigenstates
$M={\rm a}_0^0,\sigma,f_0,\zeta,\pi^0,\eta,\eta^\prime$, and 2 for $M={\rm
a}_0^+,K_0^{\ast +},K_0^{\ast 0},\pi^+,K^+,K^0$. Meson masses are then
given by the equations
\begin{equation}
G_M(-m_M^2)\ =\ 0 \ .
\end{equation}
In addition, physical states have to be normalized through
\begin{equation}
\tilde{\phi}_M(p)=Z_M^{-1/2}\ \phi_M(p)\ ,
\end{equation}
where
\begin{equation}
\label{zr}
Z_M^{-1}=\frac{dG_M(p)}{dp^2}\bigg\vert_{p^2=-m_M^2} \ .
\end{equation}

The full expressions for the one-loop functions $G_M(q)$ are quoted in the
Appendix. They can be written in terms of the coupling constants $G$ and
$H$, the mean field values $\bar S_{u,s}$, and quark loop functions that
prove to be ultraviolet convergent, owing to the asymptotic behavior of the
nonlocal form factors. For the pseudoscalar meson sector, the $\pi$ and
$K$ mesons decouple, while the $I=0$ states get mixed. Since the
corresponding mixing angles are momentum-dependent functions, it is
necessary to introduce two mixing angles $\theta_\eta$ and
$\theta_{\eta'}$, defined at $p^2 = -m_\eta^2$ and $p^2 = -m_{\eta'}^2$
respectively, see Eq.~(\ref{angles}). In the case of the scalar meson
sector, the ${\rm a}_0$ and $K_0^\ast$ mesons decouple, while the $\zeta$,
$\sigma_0$, and $\sigma_8$ fields get mixed by a $3\times 3$ matrix, see
Eq.~(\ref{mtxsc}).

One can also calculate the weak decay constants of pseudoscalar mesons.
These are given by the matrix elements of the axial currents $ A_\mu^a$
between the vacuum and the physical meson states,
\begin{equation}
\label{fpiab}
\imath f_{ab}(p^2) \; p_\mu=\langle 0 \vert A_\mu^a(0) \vert \delta\pi_b(p)
\rangle\ .
\end{equation}
The matrix elements can be calculated from the expansion of the Euclidean
effective action in the presence of external axial currents,
\begin{equation}
\label{der}
\langle 0 \vert A_\mu^a(0) \vert \delta\pi_b(p) \rangle = \frac{\delta^2
S_E}{\delta A_\mu^a \delta \pi_b(p)}\bigg\vert_{A_\mu^a=\delta\pi_b=0} \ .
\end{equation}
It is important to notice that, owing to nonlocality, the axial currents
have to be introduced not only into the covariant derivative in the
Euclidean action, but also in the fermion fields entering the nonlocal
currents, through the replacements~\cite{BB95,Scarpettini:2003fj}
\begin{eqnarray}
\psi\left(x-\frac{z}{2}\right) &\longrightarrow& W_A\left(x,x-\frac{z}{2}\right)
\psi\left(x-\frac{z}{2}\right)\nonumber\\
\psi^\dagger\left(x+\frac{z}{2}\right) &\longrightarrow &
\psi^\dagger\left(x+\frac{z}{2}\right)W_A\left(x+\frac{z}{2},x\right)\ .
\end{eqnarray}
Here the transport function
$W_A (x,y)$ is given by
\begin{equation}
W_A (x,y)\ =\ P \mbox{
exp}\left\lbrace\frac{\imath}{2}\int_x^yds_\mu\gamma_5\lambda_aA_\mu^a(s)\right\rbrace
\ ,
\end{equation}
where $s$ runs over an arbitrary path connecting $x$ with $y$.

After a rather lengthy calculation, we find that the relevant term in
the expansion of the Euclidean action can be written as
\begin{equation}
S_E^{[A,\phi]}\ = \ \int \frac{d^4 p}{(2\pi)^4} \frac{d^4
p^\prime}{(2\pi)^4}\ \sum_{i,j=1}^3 A_{\mu\, ij}(p)\
\delta\pi_{ji}(p^\prime)\ G^\mu_{ij}(p,p^\prime) \ ,
\end{equation}
where we have defined $A_\mu = \lambda_a A_\mu^a/\sqrt{2}$, $\delta\pi =
\lambda_a \delta\pi_a/\sqrt{2}$. The functions $G^\mu_{ij}(p,p^\prime)$
are found to satisfy the relation
\begin{equation}
p_\mu\, G^\mu_{ij}(p,p^\prime)\ =\ -\,i\,\delta^{(4)} (p+p^\prime)\,
F_{ij}(p^2)\ ,
\end{equation}
where
\begin{eqnarray}
\label{Fij}
F_{ij}(p^2) & = & 2 N_c  \int \frac{d^4 q}{(2\pi)^4}\; \left[g(q^+)+g(q^-)-2g(q)\right]
\, Z(q) \left[\frac{M_i(q)}{q^2 + M^2_i(q)}\, + \,
\frac{M_j(q)}{q^2 + M^2_j(q)}\right]
\nonumber\\
& & -\,2 N_c \int \frac{d^4q}{(2\pi)^4}\;
(\bar\sigma_i+\bar\sigma_j)\,
\left[g(q^+)+g(q^-)-2g(q)\right]\, g(q)
\nonumber\\
& & \times\,
\frac{Z(q^+)}{M^2_i(q^+)+q^{+2}}\;\frac{Z(q^-)}{M^2_j(q^-)+q^{-2}}
\;\left[ (q^+\cdot q^-) + M_i(q^+)M_j(q^-) \right]
\nonumber\\
& & +\,4 N_c \int \frac{d^4 q}{(2\pi)^4}\; g(q)\;
\frac{\left[M_i(q^+)q^- - M_j(q^-)q^+\right]\cdot
\left[Z(q^-)q^+-Z(q^+)q^-\right]}{[q^{+2}+M^2_i(q^+)]\,[q^{-2}+M^2_j(q^-)]} \ \
,
\end{eqnarray}
with $q^\pm = q \pm p/2$. It is worth pointing out that the functions
$F_{ij}$ (and, therefore, the weak decay constants) are given by the
longitudinal component of $G_{ij}^\mu(p,p')$, which does not depend on the
arbitrary path chosen in the transport functions $W_A(x,y)$.

{}From the above expressions, the weak decay constants for $\pi$ and $K$
mesons in the isospin limit are given by
\begin{eqnarray}
f_\pi &=& \frac{Z_\pi^{1/2}}{m_\pi^2}\; F_{uu}(p^2)\bigg|_{p^2=-m_\pi^2}\ ,
\nonumber\\
f_K &=& \frac{Z_K^{1/2}}{m_K^2}\; F_{us}(p^2)\bigg|_{p^2=-m_K^2} \ .
\end{eqnarray}
For the $\eta-\eta^{\prime}$ sector, the functions $f_{ab}(p^2)$ defined in
Eq.~(\ref{fpiab}) are related to $F_{ij}(p^2)$ through
\begin{eqnarray}
f_{00}(p^2) &=& \frac{1}{3}\left[ 2F_{uu}(p^2)+F_{ss}(p^2)\right]\ ,\nonumber\\
f_{88}(p^2) &=& \frac{1}{3}\left[ F_{uu}(p^2)+2F_{ss}(p^2)\right]\ ,\nonumber\\
f_{08}(p^2) &=& \frac{\sqrt{2}}{3}\left[ F_{uu}(p^2)- F_{ss}(p^2)\right] \ .
\end{eqnarray}
These can be translated to the mass eigenstate basis through the mixing
angles in Eq.~(\ref{angles}). Thus one defines
\begin{eqnarray}
f_{\eta}^a &=& \frac{Z_\eta^{1/2}}{m_\eta^2} \left[f_{a8}(p^2)\cos
\theta_\eta - f_{a0}(p^2) \sin\theta_\eta\right]\bigg|_{p^2=-m_\eta^2} \ ,
\qquad a=0,8 \ ,
\nonumber \\
f_{\eta^\prime}^a &=&
\frac{Z_{\eta^\prime}^{1/2}}{m_{\eta^\prime}^2}\left[f_{a8}(p^2)
\sin\theta_{\eta^\prime} + f_{a0}(p^2)
\cos\theta_{\eta^\prime}\right]\bigg|_{p^2=-m_{\eta^\prime}^2} \ , \qquad
a=0,8 \ .
\end{eqnarray}
In order to compare with phenomenological determinations, it is convenient
to consider an alternative parametrization in terms of two decay constants
$f_0$, $f_8$ and two mixing angles $\theta_0$, $\theta_8$~\cite{L97,F00}.
Both parametrizations are related by
\begin{equation}
\left( \begin{array}{cc} f_{\eta}^8 & f_{\eta}^0 \\
f_{\eta^\prime}^8  & f_{\eta^\prime}^0 \end{array} \right) =
\left( \begin{array}{cc} f_8\cos\theta_8 & -f_0\sin\theta_0 \\
f_8\sin\theta_8  & f_0\cos\theta_0 \end{array} \right) \ .
\end{equation}

\subsection{\sc Model parameters and form factors}

The model includes five parameters, namely the current quark masses
$m_{u,s}$ and the coupling constants $G$, $H$, and $\kappa$. In addition,
one has to specify the form factors $f(z)$ and $g(z)$ entering the
nonlocal fermion currents. Here, following Ref.~\cite{Noguera:2008}, we
will consider two parameter sets, corresponding to two different
functional forms for $f(z)$ and $g(z)$. The first one corresponds to the
often-used exponential forms
\begin{equation}
g(p)= \mbox{exp}\left(-p^{2}/\Lambda_{0}^{2}\right) \ ,
\qquad f(p)= \mbox{exp}\left(-p^{2}/\Lambda_{1}^{2}\right)\ ,
\label{ff1}
\end{equation}
which guarantee a fast ultraviolet convergence of the loop integrals. Note
that the range (in momentum space) of the nonlocality in each channel is
determined by the parameters $\Lambda_0$ and $\Lambda_1$, respectively.
The second set of form factors considered here is
\begin{eqnarray}
g(p)  = \frac{1+\alpha_z}{1+\alpha_z\ f_z(p)} \frac{\alpha_m \ f_m (p) -m\
\alpha_z f_z(p)} {\alpha_m - m \ \alpha_z } \ , \qquad f(p)  = \frac{ 1+
\alpha_z}{1+\alpha_z \ f_z(p)} f_z(p)\ ,
\label{ff2}
\end{eqnarray}
where
\begin{equation}
f_{m}(p) = \left[ 1+ \left( p^{2}/\Lambda_{0}^{2}\right)^{3/2} \right]^{-1}
\ , \qquad
f_{z}(p) = \left[ 1+ \left( p^{2}/\Lambda_{1}^{2}\right) \right]^{-5/2}.
\label{faux2}
\end{equation}
As shown in Ref.~\cite{Noguera:2008}, for the SU(2) version of the model
these functional forms can very well reproduce the momentum dependence of
mass and wave function renormalization obtained in lattice calculations.

\begin{figure}[h]
\centering
\subfloat{\includegraphics[scale=0.24]{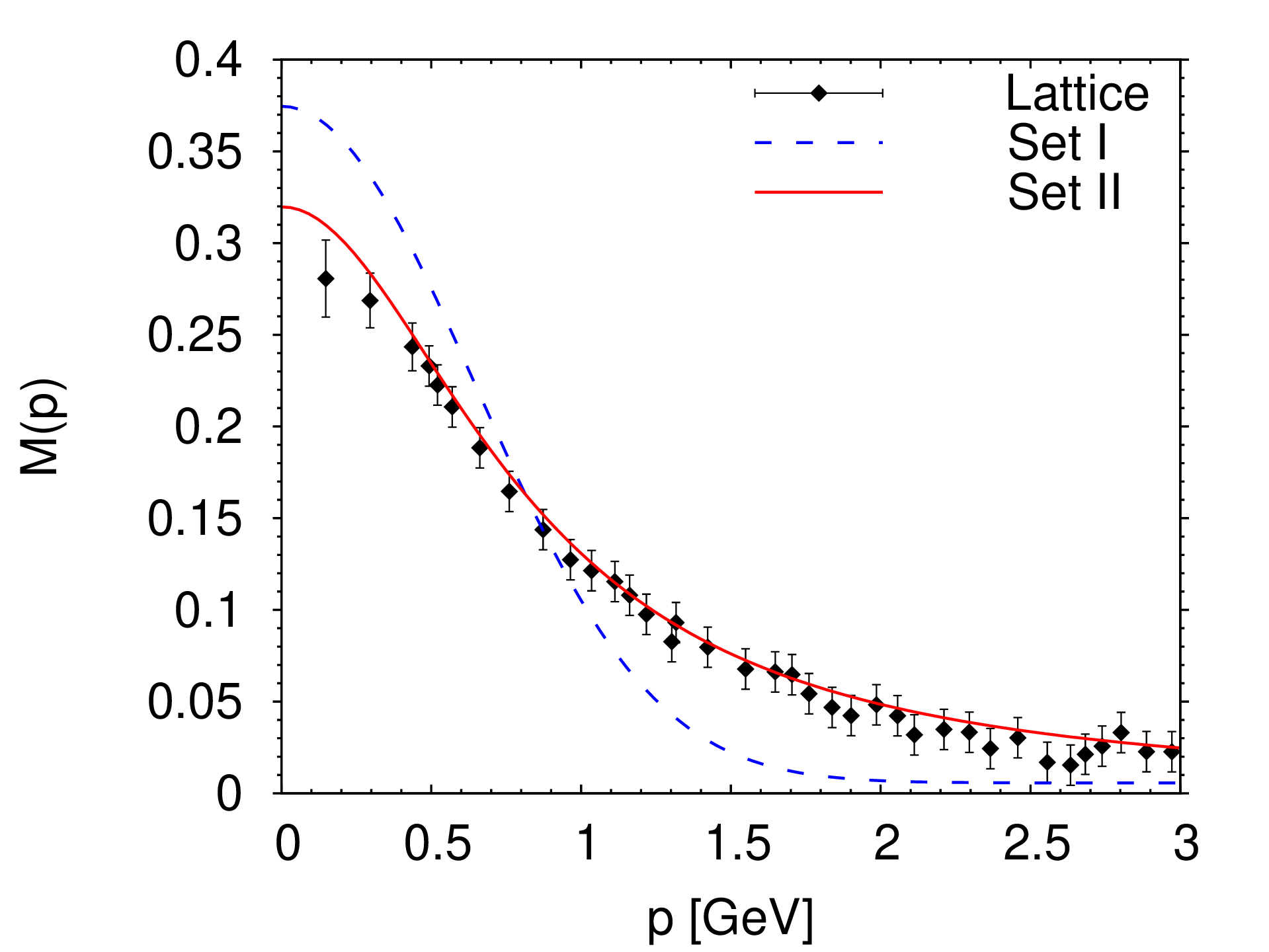}}
\subfloat{\includegraphics[scale=0.24]{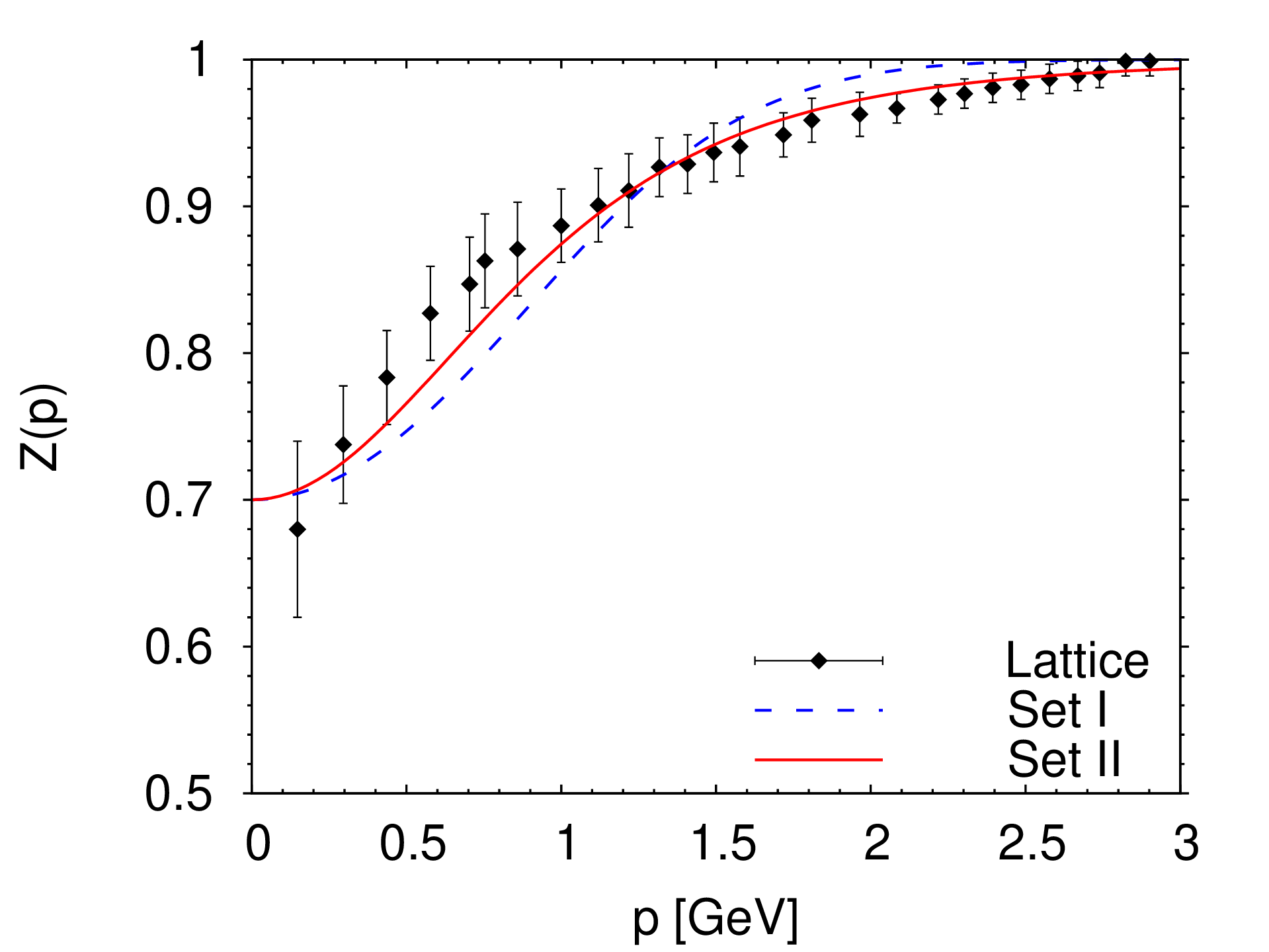}}
\caption{Mass and WFR as functions of the momentum for our parameterizations
sets I and II, in comparison with lattice results from
Ref.~\cite{Parappilly:2005ei}.}
\label{fig:1}
\end{figure}

Given the form factor functions, one can fix the model parameters so as to
reproduce the observed meson phenomenology. To the above-mentioned
parameters $m_{u,s}$, $G$, $H$, and $\kappa$ one has to add the cutoffs
$\Lambda_0$ and $\Lambda_1$, introduced through the form factors. Here we
have chosen to take as input value the light quark mass $m_u$, while the
remaining six parameters are determined by fixing the value of the quark WFR
at momentum zero, $Z(0)=0.7$ (as dictated by lattice QCD estimations), and
by requiring that the model reproduces the empirical values of five physical
quantities. These are the masses of the pseudoscalar mesons $\pi$, $K$ and
$\eta^{\prime}$, the pion weak decay constant $f_{\pi}$ and the light quark
condensate $\langle \overline{u} u \rangle$. In Table~I we quote the
numerical results for the model parameters that we have obtained for the
above-described form factor functions. In what follows, the parameter sets
corresponding to the form factors in Eqs.~(\ref{ff1}) and
(\ref{ff2}-\ref{faux2}) will be referred to as set I and set II,
respectively. As expected from the ansatz chosen for the form factors, for
set II the momentum-dependent mass and WFR in the light quark propagators
are able to fit adequately the results obtained in lattice QCD calculations.
This is shown in Fig.~\ref{fig:1}, where we plot the curves obtained for the
functions $M(p)$ and $Z(p)$, together with $N_f=2+1$ lattice data taken from
Ref.~\cite{Parappilly:2005ei}. For comparison we also quote the results
corresponding to set I.
\begin{table} [h]
\begin{center}
\begin{tabular}{c c c}
\hline \hline
 & \ \ Set I \ \ & \ \ Set II \ \ \\
\hline \hline
$m_u$ [MeV] & 5.7  & 2.5 \\
$m_s$ [MeV] & 136  & 63.9 \\
$G\Lambda_0^2$  & 23.64  & 15.55  \\
$-H\Lambda_0^5$ & 526 & 241 \\
$\kappa$ [GeV] & 4.36 & 8.08 \\
$\Lambda_0$ [GeV] & 0.814  & 0.824 \\
$\Lambda_1$ [GeV] & 1.032  & 1.550  \\
\hline
\end{tabular}
\caption{\small{Model parameters for the form factors in Eqs.~(\ref{ff1})
(set I) and (\ref{ff2}-\ref{faux2}) (set II).}}
\end{center}
\label{TableI:param}
\end{table}

\subsection{\sc Meson phenomenology}
\label{sect:mesons}

Once the parameters have been determined, we can calculate the values of
several meson properties for the scalar and pseudoscalar sectors. Our
numerical results for sets I and II are summarized in Table~II, together
with the corresponding phenomenological estimates. The quantities marked
with an asterisk are those that have been chosen as input values. In
general, it is seen that the meson masses, mixing angles, and weak decay
constants predicted by the model are in reasonable agreement with
phenomenological expectations. Moreover, the results for set I do not differ
significantly from those obtained in Ref.~\cite{Contrera:2009hk} for a
nlPNJL model with a Gaussian form factor $g(p)$ and no WFR. Regarding the
scalar meson sector, a new ingredient with respect to the model with no WFR
is the presence of the additional field $\zeta$, which mixes with the $I=0$
fields $\sigma_0$ and $\sigma_8$. The mass of the physical particles can be
obtained by determining the zeros of the functions
$G_{\zeta,\sigma,f_0}(p^2)$ arising from the diagonalization of the $3\times
3$ matrix in Eq.~(\ref{mtxsc}) (see the Appendix). From the corresponding
numerical calculation it is seen that one of these functions is positive
definite for the momentum range described by our models, which reflects that
the eigenstate associated with $\zeta$ does not correspond to a physical
particle. For the remaining two states, which can be interpreted as the
$f_0(500)$ (or $\sigma$) and $f_0(980)$ scalar mesons quoted by the Particle
Data Group (PDG)~\cite{Beringer:1900zz}, we obtain masses of about 550 and
1200~MeV. In fact, in the case of the $f_0$ meson it happens that the loop
integrals in Eq.~(\ref{cij}) become divergent, and need some regularization
prescription. This occurs since $p^2$ exceeds a threshold above which both
effective quarks can be simultaneously on shell, which can be interpreted as
the possibility of a decay of the meson into two massive quarks. The
integrals can be properly defined e.g.~following the prescription in
Ref.~\cite{Scarpettini:2003fj}. Since the threshold lies at about 1 GeV, we
have estimated the mass value for the $f_0$ meson by an extrapolation from
the momentum region in which the integrals are well defined. Following a
similar procedure, the masses of $K_0^\ast$ charged and neutral mesons are
found to be $\simeq 1300$~MeV; thus, these particles can be identified with
the $K_0^\ast(1430)$ mesons quoted by the PDG ($m_{K_0^\ast} = 1425\pm
50$~MeV)~\cite{Beringer:1900zz}. Our models do not seem to include the light
strange scalar mesons $\kappa$, which indeed still need confirmation and
thus have been omitted from PDG particle summary tables.

\begin{table} [H]
\label{resultados}
\begin{center}
\begin{tabular}{c|c|c|c}
\hline \hline
 & Set I & Set II & Empirical \\
\hline \hline
$\bar\sigma_u$ [MeV] & 529 & 454 & - \\
$\bar\sigma_s$ [MeV] & 702 & 663 & - \\
$\bar\zeta / \kappa$ & $-0.429$ & $-0.429$ & - \\ 
$-\langle \overline{u} u \rangle ^{1/3}$ [MeV] $^\ast$ & 240 & 320 & - \\
$-\langle \overline{s} s \rangle ^{1/3}$ [MeV] & 198 & 343 & - \\
\hline
$m_{\pi}$ [MeV] $^\ast$ &  139   & 139 & 139 \\
$m_{K}$ [MeV] $^\ast$ & 495 & 495 & 495 \\
$m_{\eta}$ [MeV] & 527 & 537 & 547 \\
$m_{\eta^{\prime}}$ [MeV] $^\ast$ & 958 & 958 & 958 \\
$m_{{\rm a}_0}$ [MeV] & 936 & 916 & 980 \\
$m_{K_0^\ast}$ [MeV] & 1300 & 1300 & 1425 \\
$m_{\sigma}$ [MeV] & 599 & 537 & 400 - 550 \\
$m_{f_0}$ [MeV] & 1300 & 1200 & 990 \\
\hline
$f_{\pi}$ [MeV] $^\ast$ & 92.4 & 92.4 & 92.4 \\
$f_K/f_{\pi}$& 1.17 & 1.16 & 1.22 \\
$f_{\eta}^0/f_{\pi}$ & 0.17 & 0.14 & $(0.11$ - 0.507) \\
$f_{\eta}^8/f_{\pi}$ & 1.12 & 1.12 & $(1.17$ - 1.22) \\
$f_{\eta^{\prime}}^0/f_{\pi}$ & 1.09 & 1.43 & $(0.98$ - 1.16) \\
$f_{\eta^{\prime}}^8/f_{\pi}$ & $-0.48$ & $-0.42$ & $-(0.42$ -\ 0.46)\ \\
\hline
$\theta_{\eta}$ & $-2.95^\circ$ & $-1.01^\circ$  & - \\
$\theta_{\eta^{\prime}}$ & $-41.62^\circ$ & $-30.79^\circ$ & - \\
$\theta_{0}$ &  $-8.63^\circ$ & $-5.53^\circ$ & $-(0^\circ$ - 10$^\circ$)\ \\
$\theta_{8}$ &  $-22.94^\circ$ & $-20.67^\circ$ & $-(19^\circ$ - 22$^\circ$)\ \\
\hline
\end{tabular}
\caption{\small{Numerical results for various phenomenological quantities.
Input values are marked with an asterisk.}}
\end{center}
\end{table}

Concerning the quark masses and condensates, it is found that in the case of
set II we obtain relatively low values for $m_u$ and $m_s$, and a somewhat
large value for the light quark condensate. Similar results have previously
been obtained in Refs.~\cite{Noguera:2008} and~\cite{Hell:2011ic}, within
two- and three-flavor parameterizations respectively. As discussed in those
articles, this can be in part attributed to the fact that our fit to lattice
data for the function $Z(p)$ is based on the calculations in
Ref.~\cite{Parappilly:2005ei}, which correspond to a rather large
renormalization scale $\mu = 3$~GeV. On the other hand, for both sets I and
II we find that the quark mass ratio is $m_s/m_u\simeq 25$, which is
phenomenologically adequate. Something similar happens with the product
$-\langle \bar uu\rangle m_u$, which gives $7.9\times 10^{-5}$~GeV$^4$ for
set I and $8.2\times 10^{-5}$~GeV$^4$ for set II: these values are in
agreement with the scale-independent result obtained from the
Gell-Mann-Oakes-Renner relation at the leading order in the chiral
expansion, namely $-\langle \bar uu\rangle m_u = f_\pi^2 m_\pi^2/2 \simeq
8.3\times 10^{-5}$~GeV$^4$.

\section{\sc Nonzero temperature}

\subsection{\sc Polyakov loop}

As stated in the previous section, the effective action of the model
includes the interaction of quarks with color gauge fields through the
covariant derivative in the fermion kinetic term. This coupling will be
treated at the mean field level, considering that quarks move on a constant
background field $\phi = A_4 = i A_0 = i g\,\delta_{\mu 0}\, G^\mu_a
\lambda^a/2$, where $G^\mu_a$ are the SU(3) color gauge fields. Then the
traced Polyakov loop, which in the infinite quark mass limit can be taken as
order parameter of confinement, is given by $\Phi=\frac{1}{3} {\rm Tr}\,
\exp( i \phi/T)$. We will work in the so-called Polyakov gauge, in which the
matrix $\phi$ is given a diagonal representation $\phi = \phi_3 \lambda_3 +
\phi_8 \lambda_8$. Owing to the charge conjugation properties of the QCD
Lagrangian~\cite{Dumitru:2005ng}, the mean field traced Polyakov loop field
$\Phi$ is expected to be a real quantity. Assuming that $\phi_3$ and
$\phi_8$ are real valued~\cite{Roessner:2006xn}, this implies $\phi_8 = 0$,
$\Phi = [ 2 \cos(\phi_3/T) + 1 ]/3$.

The effective gauge field self-interactions are given by the Polyakov-loop
potential ${\cal U}\,[A(x)]$. At finite temperature $T$, it is normal to take
for this potential a functional form based on properties of pure gauge QCD.
One possible ansatz is that based on the logarithmic expression of the Haar
measure associated with the SU(3) color group integration. The corresponding
potential is given by~\cite{Roessner:2006xn}
\begin{equation}
\frac{{\cal{U}}_{\rm log}(\Phi ,T)}{T^4} \ =
\ -\,\frac{1}{2}\, a(T)\,\Phi^2 \;+
\;b(T)\, \log\left(1 - 6\, \Phi^2 + 8\, \Phi^3
- 3\, \Phi^4 \right) \ ,
\label{ulog}
\end{equation}
where
\begin{equation}
a(T) = a_0 +a_1 \left(\dfrac{T_0}{T}\right) + a_2\left(\dfrac{T_0}{T}\right)^2
\ ,
\qquad
b(T) = b_3\left(\dfrac{T_0}{T}\right)^3 \ .
\label{log}
\end{equation}
The parameters can be fitted to pure gauge lattice QCD data so as to
properly reproduce the corresponding equation of state and Polyakov loop
behavior. This leads to~\cite{Roessner:2006xn}
\begin{equation}
a_0 = 3.51\ ,\qquad a_1 = -2.47\ ,\qquad a_2 = 15.2\ ,\qquad b_3 = -1.75\ .
\end{equation}
The values of $a_i$ and $b_i$ are constrained by the condition of reaching
the Stefan-Boltzmann limit at $T \rightarrow \infty$ and by imposing the
presence of a first-order phase transition at $T_0$, which is a further
parameter of the model. In the absence of dynamical quarks, from lattice
calculations one expects a deconfinement temperature $T_0 = 270$~MeV.
However, it has been argued that in the presence of light dynamical quarks
this temperature scale should be adequately reduced to about 210 and 190~MeV
for the case of two and three flavors, respectively, with an uncertainty of
about 30 MeV~\cite{Schaefer:2007pw}. Besides the logarithmic function in
Eq.~(\ref{ulog}), other forms for the Polyakov-loop potential can be found
in the literature. In the following some of them will be considered, and the
effect of the parameter $T_0$ on the phase transitions will be analyzed.

\subsection{\sc Thermodynamics}

To investigate the phase transitions and the temperature dependence of
thermodynamical quantities within our model, we consider the thermodynamical
potential per unit volume at the mean field level. We will proceed by using
the standard Matsubara formalism, following the same prescriptions as in
previous works, see e.g.~Refs.~\cite{Scarpettini:2003fj,GomezDumm:2001fz}.
In this way we obtain
\begin{equation}
\Omega^{\rm MFA} \ = \ \Omega^{\rm reg} + \Omega^{\rm free} +
\mathcal{U}(\Phi,T) + \Omega_0 \ ,
\end{equation}
where
\begin{eqnarray}
\Omega^{\rm reg} &=& -2\, T\; \sum_{n=-\infty}^{\infty}
\sum_{c,f} \int \dfrac{d^3p}{(2\pi)^3}\; \log
\left[\dfrac{p_{nc}^2 + M_f^2(p_{nc})}{Z^2(p_{nc})\, (p_{nc}^2 + m^2_f)} \right]
\nonumber \\
 & &  - \left(\bar\zeta\, \bar R + \dfrac{G}{2}\,\bar R^2 +
\dfrac{H}{4}\,\bar S_u\, \bar S_d \, \bar S_s \right)
- \dfrac{1}{2}\, \sum_f \left( \bar \sigma_f \bar S_f
+ \dfrac{G}{2}\, \bar S_f^2 \right)\ ,
\nonumber \\
\Omega^{\rm free} &=& -2\, T \sum_{c,f} \sum_{s=\pm 1} \int
\dfrac{d^3p}{(2\pi)^3}\; {\rm Re}\,\log
\left[1+\exp\left(-\frac{\epsilon_{fp} + \imath\, s\,\phi_c}{T}
\right)\right]\ .
\end{eqnarray}
Here we have defined $p_{nc}^2 = [(2n+1)\pi T+\phi_c]^2 + \vec p^{\;2}$,
$\epsilon_{fp}=\sqrt{\vec p^{\;2} + m_f^2}$. The sums over color and flavor
indices run over $c=r,g,b$ and $f=u,d,s$, respectively, and the color
background fields are $\phi_r = - \phi_g = \phi_3$, $\phi_b = 0$. The term
$\Omega_0$ is just a constant that sets the value of the thermodynamical
potential at $T=0$.

Now, from the thermodynamic potential we can calculate various thermodynamic
quantities such as the energy and entropy densities, which are given by
\begin{equation}
\varepsilon = \Omega + T s \ , \qquad
s = -\dfrac{\partial \Omega}{\partial T} \ .
\end{equation}
We are also interested in the behavior of the quark condensates and the
corresponding chiral susceptibilities, defined by
\begin{equation}
\langle \bar q q \rangle = \dfrac{\partial\Omega}{\partial m_q} \ ,
\qquad \chi_q =  \dfrac{\partial \langle \bar q q \rangle
}{\partial m_q} =  \dfrac{\partial^2\Omega}{\partial m_q^2} \ ,
\qquad q = u,d,s\ .
\label{chsus}
\end{equation}
For large temperatures, the behavior of the regularized quark condensates is
dominated by the free contribution, which grows with $T$ as $\langle \bar q
q \rangle \sim - m_q T^2$. Therefore, in order to analyze the chiral
restoration transition it is normal to define a subtracted chiral condensate
\begin{equation}
\langle \bar q q\rangle_{\rm sub} \ = \ \dfrac{\langle \bar u u \rangle \,
-\, \frac{m_u}{m_s} \, \langle \bar s s\rangle}{\langle \bar u u \rangle_{0} \,
- \, \frac{m_u}{m_s} \, \langle \bar s s \rangle_{0}} \ ,
\end{equation}
where we have also introduced a normalization factor given by the values of
the chiral condensates at zero temperature.

\subsection{\sc Numerical results}

\begin{figure}[ht]
\begin{center}
\includegraphics[scale=0.29]{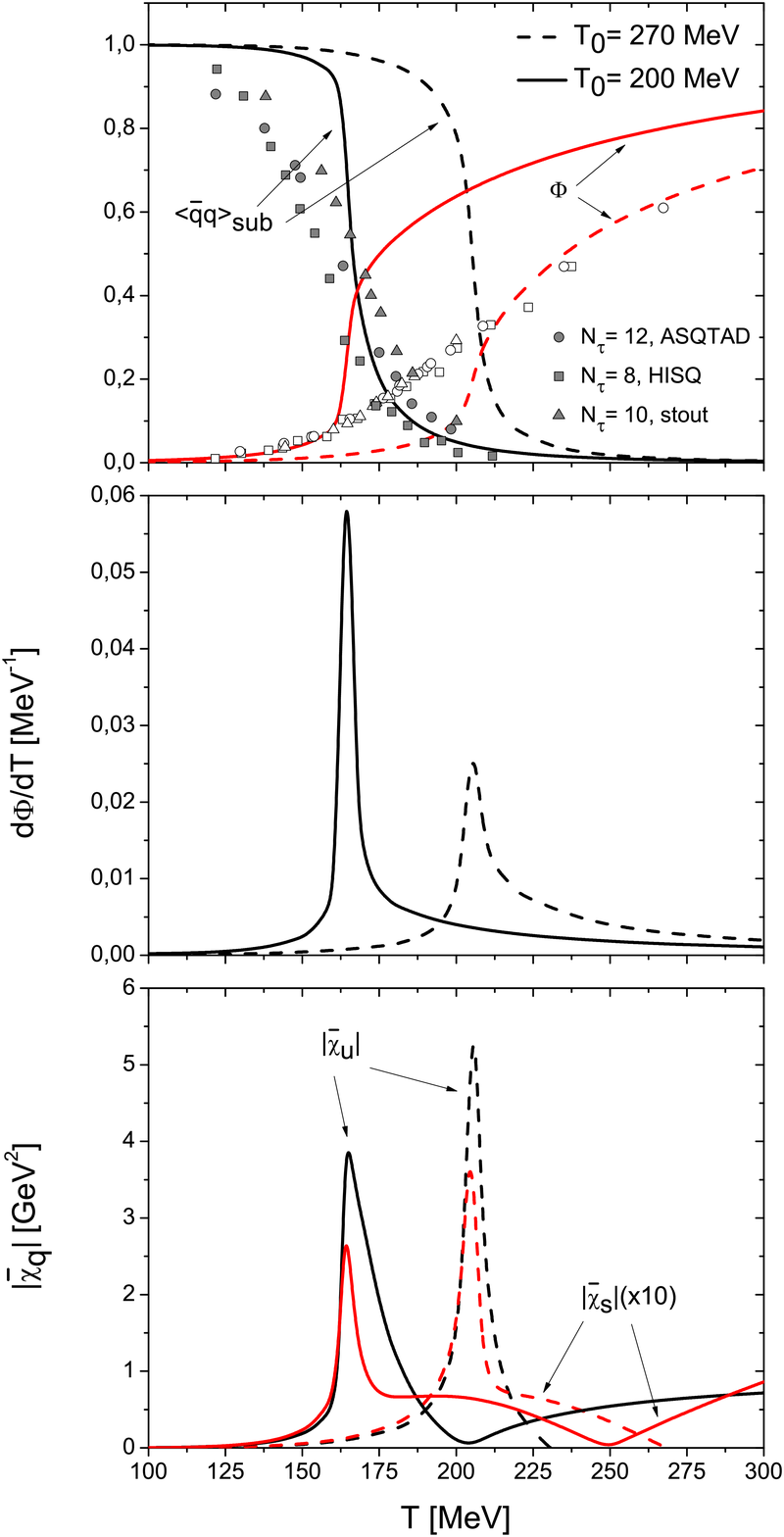}
\vspace*{-25pt}
\end{center}
\caption{Subtracted chiral condensate, Polyakov loop, chiral
susceptibilities, and Polyakov loop susceptibility $d\Phi/dT$ as functions of
the temperature. Solid (dashed) curves correspond to parameter set II, for a
logarithmic Polyakov loop potential with $T_0 = 200$ (270)~MeV. Triangles,
circles, and squares stand for lattice QCD results from
Refs.~\cite{Borsanyi:2010bp,Bazavov:2010sb}.} \label{fig:2}
\end{figure}

We present here our numerical results for the quantities defined in the
previous section, considering different form factors and Polyakov loop
potentials. Let us start by taking into account the parameterization set II,
based on lattice QCD results for the quark propagators, and the logarithmic
Polyakov-loop potential in Eq.~(\ref{ulog}). In Fig.~\ref{fig:2} we quote
the corresponding behavior of the subtracted chiral condensate, the traced
Polyakov loop $\Phi$ and the associated susceptibilities as functions of the
temperature. In the upper panel we show the results for the subtracted
chiral condensate $\langle\bar qq\rangle_{\rm sub}$ and the traced Polyakov
loop $\Phi$, for $T_0 = 270$ and 200~MeV (dashed and solid curves,
respectively). As stated, $T_0 = 270$~MeV is the deconfinement transition
temperature obtained from lattice calculations in pure gauge QCD, while we
have taken $T_0 = 200$~MeV as a reference temperature arising from the
corresponding rescaling in the presence of dynamical
quarks~\cite{Schaefer:2007pw}. For comparison we also include lattice QCD
data taken from Refs.~\cite{Borsanyi:2010bp,Bazavov:2010sb}. As expected, it
is found that when the temperature is increased the system undergoes both
the chiral restoration and deconfinement transitions, which proceed as
smooth crossovers for the considered values of $T_0$. In the central and
lower panels of Fig.~\ref{fig:2} we show the curves for the Polyakov loop
susceptibility ---defined as $d\Phi/dT$--- and the chiral susceptibilities
$\chi_{u,s}$ [given by Eq.~(\ref{chsus})] as functions of the temperature.
As usual, we take the position of the peaks to define the corresponding
transition critical temperatures. From the figure it is seen that the curves
get steeper for lower values of $T_0$; in fact, first order phase
transitions are found for $T_0 \lesssim 185$~MeV. In addition, in the curves
for $\chi_s$ it is possible to identify a second, broad peak that allows us
to define an approximate critical temperature for the restoration of the
full SU(3) chiral symmetry. For clarity we have plotted in the graphs the
subtracted susceptibilities $\bar \chi_{q} \equiv \chi_{q} - \chi_q(T=0)$.

It is seen that both the SU(2) chiral restoration and deconfinement
transitions occur essentially at the same critical temperatures, in
agreement with lattice QCD results. The numerical values from the plots in
Fig.~\ref{fig:2} are $T_c \simeq 200$~MeV and $T_c \simeq 165$~MeV for $T_0
= 270$ and 200 MeV, respectively, while lattice QCD analyses lead to a
transition temperature of about 160
MeV~\cite{Borsanyi:2010bp,Bazavov:2010sb}. Thus, the agreement with lattice
QCD data favors the suggested rescaling of the reference temperature $T_0$
from the pure gauge transition temperature towards values around 200~MeV.

\begin{figure}[h]
\begin{center}
\includegraphics[scale=0.27]{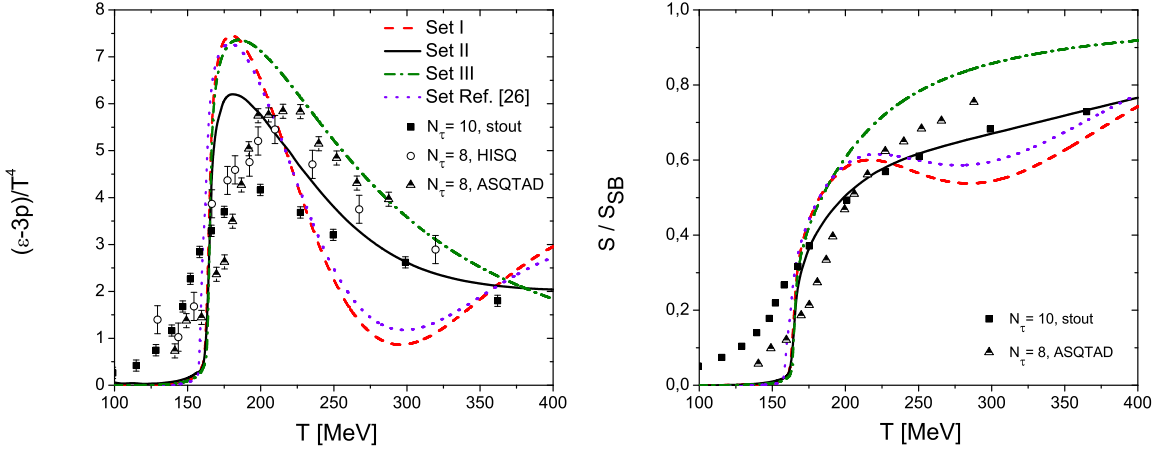}
\end{center}
\vspace*{-25pt} \caption{Normalized interaction energy (left) and entropy
density (right) as functions of the temperature, for different model
parameterizations. Curves correspond to nlPNJL models with logarithmic
Polyakov loop potentials, with $T_0 = 200$~MeV. Squares, circles, and
triangles stand for lattice data from Refs.~\cite{Bazavov:2010sb},
\cite{Bazavov:2009zn} and \cite{Borsanyi:2010cj}, respectively.}
\label{fig:3}
\end{figure}

The above results, which correspond to the lattice QCD-inspired form factors
of our parameterization set II, are qualitatively similar to those obtained
for the case of set I, based on Gaussian form factors. In order to compare
the features of parameterizations I and II, it is useful to consider other
thermodynamical quantities, such as the interaction energy and the entropy.
The corresponding curves are shown in Fig.~\ref{fig:3}, where we plot the
normalized interaction energy $(\varepsilon - 3p)/T^4$ (left) and the
normalized entropy density $s/s_{SB}$ (right), where $s_{SB}$
stands for the entropy density Stefan-Boltzmann limit. Dashed and solid
curves correspond to parameterization sets I and II, respectively, for the
logarithmic Polyakov loop potential in Eq.~(\ref{ulog}) with $T_0 =
200$~MeV. We have included, for comparison, three sets of lattice data,
taken from Refs.~\cite{Bazavov:2010sb,Bazavov:2009zn,Borsanyi:2010cj}. It
can be seen that for both the interaction energy and the entropy, the curves
for set I show a pronounced dip at about $T\sim 300$~MeV, which is not
observed in the case of set II, where the falloff is smooth. In order to
trace the source of this effect we have also considered a third
parameterization set III in which the form factor $g(p)$ has a Gaussian
shape as in set I, but we do not include the coupling driven by the currents
$j^r(x)$ [i.e.~there is no wave function renormalization, $Z(p)=1$]. This
parameterization has previously been considered in
Ref.~\cite{Contrera:2009hk}, where the values of model parameters can be
found (see also Ref.~\cite{Hell:2009by}). In Fig.~\ref{fig:3} it corresponds
to the dashed-dotted curve, which does not show the mentioned dip. This
indicates that the effect can be attributed to the exponential behavior of
the form factor $f(p)$ in the wave function renormalization for set~I.
Moreover, our results can also be compared with those obtained from the
parameterization considered in Ref.~\cite{Hell:2011ic}, where the form
factors are introduced so as to fit lattice results for the quark propagator
(as in our set II), but $f(p)$ is assumed to have a Gaussian shape. The
curves for the interaction energy and the entropy for this model (dotted
lines in Fig.~\ref{fig:3}) are similar to those obtained for our
parameterization set I. Thus, from the comparison with lattice data, one can
conclude that the choice of a powerlike behavior for $f(p)$, such as that
proposed in Eqs.~(\ref{ff2}-\ref{faux2}), turns out to be more adequate than
the exponential one.

Another aspect to be analyzed is the steepness of the curves in the
transition region. From both the plots in Figs.~\ref{fig:2} and \ref{fig:3},
it is seen that the transition predicted by the nlPNJL models is too sharp
in comparison with lattice estimations. In order to study the robustness of
this behavior, it is interesting to consider different forms for the
Polyakov loop potential proposed in the literature. Besides the logarithmic
form in Eq.~(\ref{ulog}), a widely used potential is that given by a
polynomic function based on a Ginzburg-Landau
ansatz~\cite{Ratti:2005jh,Scavenius:2002ru}:
\begin{eqnarray}
\frac{{\cal{U}}_{\rm poly}(\Phi ,T)}{T ^4} \ = \ -\,\frac{b_2(T)}{2}\, \Phi^2
-\,\frac{b_3}{3}\, \Phi^3 +\,\frac{b_4}{4}\, \Phi^4 \ ,
\label{upoly}
\end{eqnarray}
where
\begin{eqnarray}
b_2(T) = a_0 +a_1 \left(\dfrac{T_0}{T}\right) + a_2\left(\dfrac{T_0}{T}\right)^2
+ a_3\left(\dfrac{T_0}{T}\right)^3\ .
\label{pol}
\end{eqnarray}
Here the reference temperature $T_0$ plays the same role as in the
logarithmic potential in Eq.~(\ref{ulog}). Once again, the parameters can be
fitted to pure gauge lattice QCD results so as to reproduce the
corresponding equation of state and Polyakov loop behavior (numerical values
can be found in Ref.~\cite{Ratti:2005jh}). Another widely considered form is
the Polyakov loop potential proposed by
Fukushima~\cite{Fukushima:2003fw,Fukushima:2008wg}, which includes both a
logarithmic piece and a quadratic term with a coefficient that falls
exponentially with the temperature:
\begin{equation}
{\cal{U}}_{\rm Fuku}(\Phi ,T) \ = \ -\,b\,T\left[\, 54\, \exp (-a/T)\,
\Phi^2 \, + \,\log\left(1 - 6\, \Phi^2 + 8\, \Phi^3 - 3\, \Phi^4 \right)\,
\right]\ .
\label{ufuku}
\end{equation}
Values of dimensionful parameters $a$ and $b$ are given in
Ref.~\cite{Fukushima:2008wg} (notice that these lead to $T_c\simeq 200$~MeV,
a somewhat large transition temperature in comparison with present lattice
QCD estimations). Finally, we consider here the ``improved'' Polyakov loop
potential forms recently proposed in Ref.~\cite{Haas:2013qwp}, where the
full QCD potential ${\cal{U}}_{\rm glue}$ is related to a Yang-Mills
potential ${\cal{U}}_{\rm YM}$:
\begin{equation}
\frac{{\cal{U}}_{\rm glue}(\Phi ,t_{\rm glue})}{T ^4} \ = \
\frac{{\cal{U}}_{\rm YM}[\Phi ,t_{\rm YM}(t_{\rm glue})]}{T_{\rm YM}^4}\ ,
\label{utchica}
\end{equation}
where
\begin{equation}
t_{\rm YM}(t_{\rm glue}) \ = \ 0.57\, t_{\rm glue} \ = \
0.57 \left(\frac{T - T_c^{\rm glue}}{T_c^{\rm glue}}\right) \ .
\label{tglue}
\end{equation}
The dependence of the potential on the Polyakov loop $\Phi$ is taken from an
ansatz such as those in Eq.~(\ref{upoly}) or (\ref{ulog}), while for
$T_c^{\rm glue}$ a preferred value of 210 MeV is
obtained~\cite{Haas:2013qwp}.

\begin{figure}[h]
\begin{center}
\includegraphics[scale=0.27]{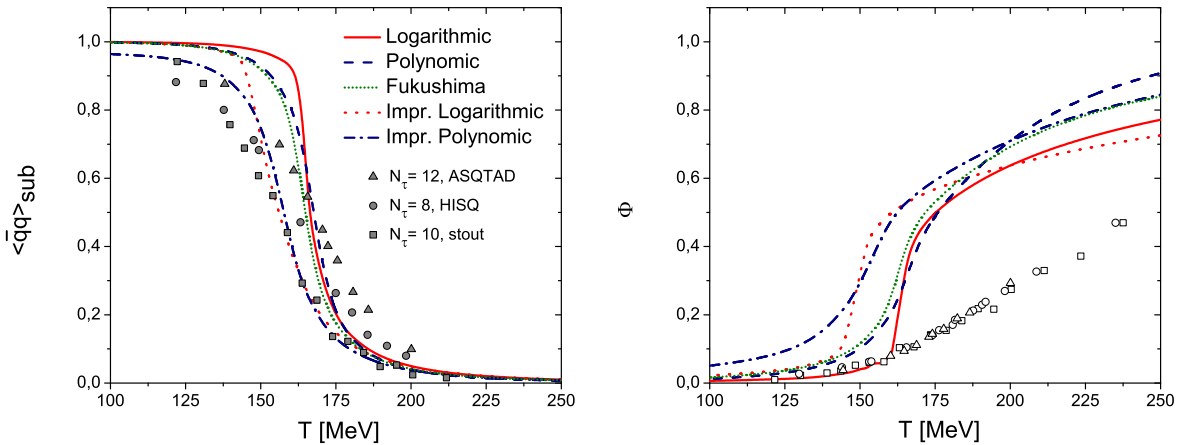}
\end{center}
\vspace*{-25pt} \caption{Subtracted chiral condensate (left) and traced
Polyakov loop (right) as functions of the temperature, for different
Polyakov loop potentials. Curves correspond to parameterization set II.
Squares, circles, and triangles stand for lattice data from
Refs.~\cite{Borsanyi:2010bp,Bazavov:2010sb}.} \label{fig:4}
\end{figure}

The transition shapes induced by these Polyakov loop potentials within our
framework are shown in Fig.~\ref{fig:4}. We have taken $T_0 = 200$~MeV for
the logarithmic and polynomic potentials, and the parameters $b$ and
$T_c^{\rm glue}$ in Eqs.~(\ref{ufuku}) and (\ref{tglue}) have been rescaled
to $(145\ {\rm MeV})^3$ and 200~MeV, respectively, in order to get critical
temperatures of about 165~MeV. From the curves for the subtracted chiral
condensate (left panel) it is seen that for the case of the polynomic and
Fukushima potentials the transition is slightly smoother than for the
logarithmic one. Moreover, the improved potentials proposed in
Ref.~\cite{Haas:2013qwp} lead to even smoother transitions, showing a
reasonable agreement with lattice QCD estimations. On the other hand, by
looking at the curves for the Polyakov loop $\Phi$ (right panel in
Fig.~\ref{fig:4}) one finds that the transition is too steep in comparison
with lattice data. This is a general feature of Polyakov NJL-like models,
both local and nonlocal, and also extends to quark-meson models. In fact, as
discussed in Refs.~\cite{Braun:2007bx,Marhauser:2008fz,Herbst:2013ufa}, the
strict comparison between our curves and lattice data for the traced
Polyakov loop has to be taken with some care, owing to the difference
between the definitions of $\Phi$ in the continuum and on the lattice. One
should expect a coincidence in the crossover temperatures, which in general
appears to be satisfied in the nlPNJL models for the potentials considered
here.

It is important to remark that in nlPNJL models one finds an entanglement
between both chiral restoration and deconfinement transitions, in agreement
with lattice QCD results. This feature is usually not observed in local PNJL
models, where both transitions appear to be typically separated by about 20
MeV, or even more (see e.g.~Refs.~\cite{Fu:2007xc,Costa:2008dp}). Something
similar happens in the region of imaginary chemical potential, where the
entanglement between both transitions occurs in a natural way within
nonlocal models~\cite{Pagura:2011rt}, while in the PNJL model it can be
obtained only after e.g.~the inclusion of an eight-quark
interaction~\cite{Sakai:2009dv}. This discrepancy with lattice QCD results
can be cured after the inclusion of an ``entangled scalar interaction'', in
which the effective four-quark coupling is a function of the traced Polyakov
loop $\Phi$~\cite{Sakai:2010rp,Sasaki:2011wu}. It is also worth noticing
that while the local PNJL in general predicts smoother transitions than the
nlPNJL, this feature should not be seen as a consequence of the nonlocality.
In fact, the enhancement of the steepness arises from the feedback between
both chiral restoration and deconfinement transitions. This is supported by
the results found in the above-mentioned ``entangled'' PNJL: by including a
$\Phi$-dependent interaction that leads to simultaneous critical
temperatures at about $175$~MeV, the transitions become steeper, just as
those obtained in nlPNJL models.

\begin{figure}[h]
\begin{center}
\includegraphics[scale=0.29]{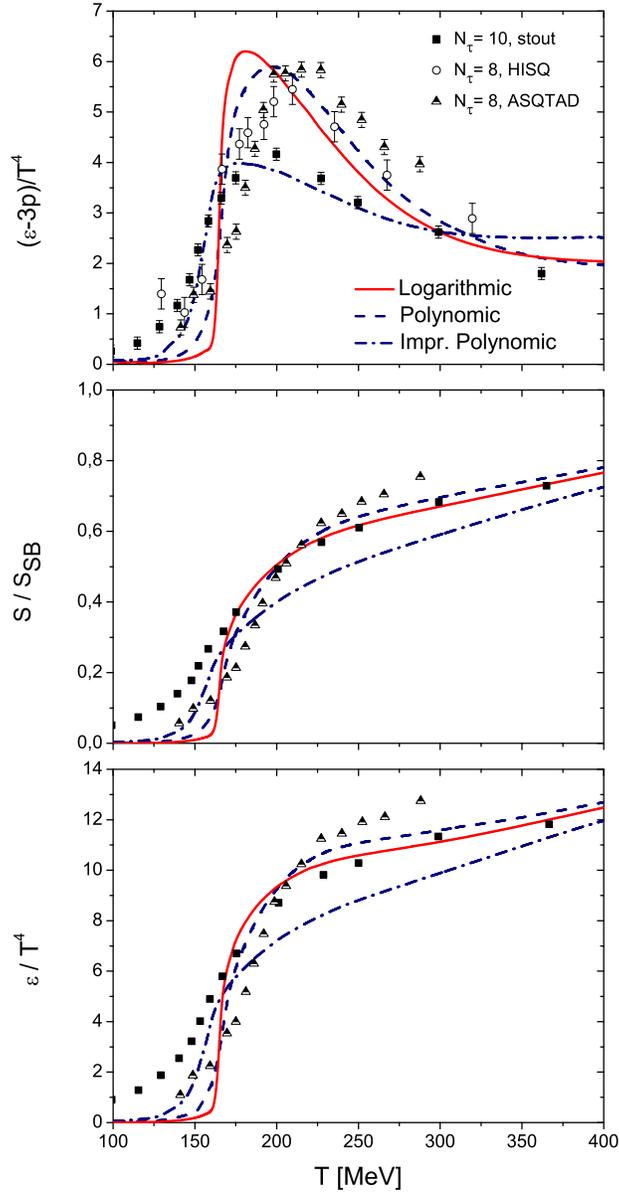}
\end{center}
\vspace*{-15pt} \caption{Normalized interaction energy, entropy density and
energy density as functions of the temperature, for parameterization set II
and three different Polyakov loop potentials. Squares, circles and triangles
stand for lattice data from
Refs.~\cite{Bazavov:2010sb,Bazavov:2009zn,Borsanyi:2010cj}.} \label{fig:5}
\end{figure}

Finally, for completeness we quote in Fig.~\ref{fig:5} our results for the
behavior of the interaction energy and the entropy and energy densities as
functions of the temperature, considering both the logarithmic and the
polynomic Polyakov loop potentials, as well as the improved polynomic
potential from Ref.~\cite{Haas:2013qwp}. The curves correspond to our
parameter set II. It is seen that the improved potential seems to be more
compatible with lattice results up to the critical temperature, while at
higher temperatures the agreement is better for the usual logarithmic and
polynomic potentials. Concerning the steepness of the transitions, it is
worth mentioning that the behavior may be softened after the inclusion of
mesonic corrections to the Euclidean action, since when the temperature is
increased the light mesons should be excited before the
quarks~\cite{Blaschke:2007np,Hell:2008cc,Hell:2009by,Radzhabov:2010dd}. The
incorporation of meson fluctuations should not modify the critical
temperatures, which for the parameters chosen here are in good agreement
with lattice estimations.

\section{\sc Summary and Conclusions}

We analyze here the features of three-flavor nlPNJL models that include a
wave function renormalization in the effective quark propagators. This
represents an extension of previous works that consider two-flavor schemes,
and three-flavor models with no quark WFR. In this framework, we obtain a
parameterization of the model that reproduces lattice QCD results for the
momentum dependence of the effective quark mass and WFR, and at the same
time leads to an acceptable phenomenological pattern for particle masses and
decay constants in both the scalar and pseudoscalar meson sectors. For
comparison we also consider a parameterization based on Gaussian form
factors, which leads to a faster convergence of quark loop integrals.
Gaussian and lattice-inspired parameterizations are called here set I and
set II, respectively. It is seen that the predictions for meson properties
are qualitatively similar in both cases, and they also agree with those
obtained previously within three-flavor models with no WFR.

As a second step we analyze the characteristics of the deconfinement and
chiral restoration transitions at finite temperature, introducing the
couplings of fermions to a background gauge field and taking the traced
Polyakov loop as order parameter for the deconfinement. In general it is
found that both transitions occur at the same critical temperature, in
agreement with lattice QCD results. This temperature turns out to be
strongly dependent on the scale parameter $T_0$ in the Polyakov loop
potential. A critical temperature of about 170~MeV, consistent with that
arising from lattice QCD calculations, is obtained for $T_0\simeq 200$~MeV,
in agreement with theoretical expectations for a model with two/three light
dynamical quarks. On the other hand, in order to distinguish between the
different parameterizations and those proposed in related works, we analyze
the temperature dependence of the interaction energy and the normalized
entropy and energy densities. For these thermodynamical quantities it is
seen that the lattice-inspired powerlike parameterization set II shows
indeed the best agreement with lattice QCD results, which supports the
consistency of our approach. We also consider various possible forms for the
Polyakov loop potential, tuning the corresponding parameters so as to obtain
critical temperatures in a range compatible with lattice QCD data. It is
seen that while the logarithmic potential Eq.~(\ref{ulog}) leads to rather
sharp transitions, for some alternative forms such as the polynomic and
Fukushima potentials [Eqs.~(\ref{upoly}) and (\ref{ufuku}), respectively]
the crossovers tend to be somewhat smoother. Moreover, the improved
potentials proposed in Ref.~\cite{Haas:2013qwp} lead to even smoother
transitions, showing a reasonable agreement with lattice QCD estimations.
The compatibility with the results for the other mentioned thermodynamical
quantities is also discussed. Finally, it is worth mentioning that the
transitions could also be softened after the incorporation of meson
fluctuations, which is presently under study in the context of our models.

\section*{Acknowledgments}

This work has been partially funded by CONICET (Argentina) under Grants No.\
PIP 00682 and No.\ PIP 02495, and by ANPCyT (Argentina) under Grant No.\
PICT-2011-0113.

\section*{\sc Appendix: analytic expressions for $G_M(p^2)$ functions}

We present here the analytic expressions for the functions $G_M(p^2)$
appearing in the quadratic expansion of the Euclidean action, see
Eq.~(\ref{spiketa}). Our calculations are in agreement with the results
reported in Ref.~\cite{Hell:2011ic}. For the $I\neq 0$ states $\pi$, ${\rm
a}_0$, $K$ and $\kappa$ we obtain
\begin{eqnarray}\label{gp}
G_{
\binom{\pi}{{\rm a}_0}}(p)&=& (G\pm\frac{H}{2}\bar S_s)^{-1}+4\, C_{uu}^{\mp}(p)\ ,
\nonumber\\
G_{
\binom{K}{\kappa}}(p)&=& (G\pm\frac{H}{2}\bar S_u)^{-1}+4\, C_{us}^{\mp}(p)\ ,
\end{eqnarray}
where the functions $C_{ij}^\mp(p)$, with $i,j=u$ or $s$, are defined as
\begin{eqnarray}
C_{ij}^{\mp}(p)&=& - \, 2\, N_c \int \frac{d^4q}{(2\pi)^4}\ g^2(q)\;
\frac{Z(q^+)}{q^{+2}+M^2_i(q^+)}\
\frac{Z(q^-)}{q^{-2}+M^2_j(q^-)}\nonumber\\
& & \times\ [ (q^+\cdot q^-) \pm M_i(q^+)M_j(q^-)]\ ,
\end{eqnarray}
with $q^\pm = q \pm p/2$. In the $I=0$ pseudoscalar sector one has a
mixing between the $\eta_0$ and $\eta_8$ fields. The masses of the
physical states $\eta$ and $\eta'$ can be obtained from the functions
\begin{equation}
G_{
\binom{\eta}{\eta^\prime}}(p) \ = \ \frac{G_{88}^-(p)+G_{00}^-(p)}{2}\mp
\sqrt{G_{80}^-(p)^2+\left(\frac{G_{88}^-(p)-G_{00}^-(p)}{2}\right)^2}\ ,
\end{equation}
where we use the definitions
\begin{eqnarray}
G_{00}^{\mp}(p)&=& \frac{4}{3}\left[
2C_{uu}^{\mp}(p)+C_{ss}^{\mp}(p)+\frac{6G\mp H\overline{S}_s\pm
4H\bar S_u}{8G^2- 4H^2\bar S_u^2\mp4HG\bar S_s}
\right]
\nonumber\\
G_{88}^{\mp}(p)&=& \frac{4}{3}\left[
2C_{ss}^{\mp}(p)+C_{uu}^{\mp}(p) +
\frac{6G\mp2H\bar S_s\mp4H\bar S_u}{8G^2-4H^2\bar S_u^2\mp4HG\bar S_s} \right]
\nonumber\\
G_{80}^{\mp}(p)&=& \frac{4}{3}\sqrt{2}\left[
C_{uu}^{\mp}(p)-C_{ss}^{\mp}(p)\pm\frac{H(\bar S_s-\bar S_u)}{8G^2-4H^2
\bar S_u^2\mp4HG\bar S_s} \right] \ .
\end{eqnarray}
The states $\eta$ and $\eta'$ are thus defined as
\begin{eqnarray}
\eta &=& \eta_8 \mbox{ cos }\theta_\eta - \eta_0 \mbox{ sen }\theta_\eta \ ,
\nonumber\\
\eta^\prime &=& \eta_8 \mbox{ sen }\theta_\eta^\prime + \eta_0 \mbox{ cos
}\theta_\eta^\prime \ ,
\end{eqnarray}
where the mixing angles $\theta_\eta$, $\theta_{\eta'}$ are given by
\begin{equation}
 \mbox{ tan }2\,\theta_{\eta,\eta^\prime}=\frac{-2G_{80}^-}{G_{88}^- -G_{00}^-}
 \bigg\vert_{p^2=-m^2_{\eta,\eta^\prime}} \ .
 \label{angles}
\end{equation}

Finally, for the $I=0$ scalar sector, the quadratic terms involving the
fields $\zeta$, $\sigma_8$ and $\sigma_0$ are mixed by the $3\times 3$
matrix
\begin{equation}
\left( \begin{array}{ccc} 4\,C^\zeta(p)+G^{-1} &
\sqrt{\frac{8}{3}}[ 2\,C_{u}^{+\zeta}(p) + C_{s}^{+\zeta}(p) ] &
\frac{4}{\sqrt{3}}[ C_{u}^{+\zeta}(p) - C_{s}^{+\zeta}(p) ] \\
\sqrt{\frac{8}{3}}[ 2\,C_{u}^{+\zeta}(p) + C_{s}^{+\zeta}(p) ] & G_{00}^{+}(p) &
G_{80}^{+}(p) \\
\frac{4}{\sqrt{3}}[ C_{u}^{+\zeta} - C_{s}^{+\zeta}(p) ] &
G_{80}^{+}(p) & G_{88}^{+}(p) \end{array} \right)\ ,
\label{mtxsc}
\end{equation}
where
\begin{eqnarray}
C^\zeta(p)&=& \dfrac{ N_c}{\kappa^2}\, \int \frac{d^4q}{(2\pi)^4}\ q^2
f^2(q) \sum_{i=1}^3
\frac{Z(q^+)}{q^{+2} + M^2_i(q^+)}\frac{Z(q^-)}{q^{-2} + M^2_i(q^-)}
\nonumber \\
& & \times \left[
q^+q^- + \dfrac{q^{+2}q^{-2}-(q^+q^-)^2}{2q^2} - M_i(q^+)M_i(q^-)\right]
\nonumber\\
C_{i}^{+\zeta}(p)&=& -\dfrac{2\, N_c}{\kappa}\,\int \frac{d^4q}{(2\pi)^4}\
g(q)\, f(q)\, \frac{Z(q^+)}{q^{+2} + M^2_i(q^+)}\ \frac{Z(q^-)}{q^{-2} + M^2_i(q^-)}
\nonumber\\
& & \times\ q\cdot \left[ q^- \, M_i(q^+)+ q^+ M_i(q^-)\right] \ ,
\label{cij}
\end{eqnarray}
with $i=u,s$. For a given value of $p^2$, we denote the eigenvalues of this
matrix by $G_\zeta(p)$, $G_\sigma(p)$ and $G_{f_0}(p)$. As stated in
Sec.~\ref{sect:mesons}, from the functions $G_\sigma(p)$ and $G_{f_0}(p)$
one can determine the masses of the $\sigma$, $f_0$ physical states [the
function $G_\zeta(p)$ turns out to be positive definite for the allowed
values of $-p^2$]. The corresponding mixing angles can be obtained in a
similar way as in the $\eta$ meson sector, now defining SO(3) rotation
matrices for the $\sigma$ and $f_0$ physical states.


\begin{thebibliography}{99}

\bibitem{All03}
  C.~R.~Allton, S.~Ejiri, S.~J.~Hands, O.~Kaczmarek, F.~Karsch, E.~Laermann and C.~Schmidt,
  Phys.\ Rev.\ D {\bf 68}, 014507 (2003);
  C.~R.~Allton, M.~Doring, S.~Ejiri, S.~J.~Hands, O.~Kaczmarek, F.~Karsch, E.~Laermann and K.~Redlich,
  Phys.\ Rev.\ D {\bf 71}, 054508 (2005).

\bibitem{Fod04}
Z.~Fodor and S.~D.~Katz, JHEP {\bf 0404}, 050 (2004); Y.~Aoki,
Z.~Fodor, S.~D.~Katz and K.~K.~Szabo, JHEP {\bf 0601}, 089 (2006).

\bibitem{Kar03}
F.~Karsch and E.~Laermann, in {\it Quark Gluon Plasma III}, edited
by R.C. Hwa and X. N. Wang (World Scientific, Singapore, 2004),
arXiv:hep-lat/0305025.

\bibitem{Blaschke:2007np}
D.~Blaschke, M.~Buballa, A.~E.~Radzhabov and M.~K.~Volkov,
Yad.\ Fiz. {\bf 71}, 2012 (2008)
[Phys.\ Atom.\ Nucl. \textbf{71}, 1981 (2008)].

\bibitem{Contrera:2007wu}
G.~A.~Contrera, D.~Gomez Dumm and N.~N.~Scoccola,
Phys.\ Lett.\ B {\bf 661}, 113 (2008).

\bibitem{Contrera:2009hk}
  G.~A.~Contrera, D.~Gomez Dumm and N.~N.~Scoccola,
  Phys.\ Rev.\ D {\bf 81}, 054005 (2010).

\bibitem{Hell:2008cc}
  T.~Hell, S.~Roessner, M.~Cristoforetti and W.~Weise,
  Phys.\ Rev.\ D {\bf 79}, 014022 (2009).

\bibitem{Hell:2009by}
  T.~Hell, S.~Rossner, M.~Cristoforetti and W.~Weise,
  Phys.\ Rev.\ D {\bf 81}, 074034 (2010).

\bibitem{Meisinger:1995ih}
  P.~N.~Meisinger and M.~C.~Ogilvie,
  Phys.\ Lett.\  B {\bf 379}, 163 (1996).

\bibitem{Fukushima:2003fw}
  K.~Fukushima,
  Phys.\ Lett.\  B {\bf 591}, 277 (2004).

\bibitem{Megias:2004hj}
  E.~Megias, E.~Ruiz Arriola and L.~L.~Salcedo,
  Phys.\ Rev.\  D {\bf 74}, 065005 (2006).

\bibitem{Ratti:2005jh}
  C.~Ratti, M.~A.~Thaler and W.~Weise,
  Phys.\ Rev.\  D {\bf 73}, 014019 (2006).

\bibitem{Roessner:2006xn}
  S.~Roessner, C.~Ratti and W.~Weise,
  Phys.\ Rev.\  D {\bf 75}, 034007 (2007).

\bibitem{Mukherjee:2006hq}
  S.~Mukherjee, M.~G.~Mustafa and R.~Ray,
  Phys.\ Rev.\  D {\bf 75}, 094015 (2007).

\bibitem{Sasaki:2006ww}
  C.~Sasaki, B.~Friman and K.~Redlich,
  Phys.\ Rev.\  D {\bf 75}, 074013 (2007).

\bibitem{Schafer:1996wv}
  T.~Schafer and E.~V.~Shuryak,
  Rev.\ Mod.\ Phys.\  {\bf 70}, 323 (1998).

\bibitem{RW94}
C.~D.~Roberts and A.~G.~Williams, Prog.\ Part.\ Nucl.\ Phys.\ {\bf
33}, 477 (1994); C.~D.~Roberts and S.~M.~Schmidt, Prog.\ Part.\
Nucl.\ Phys.\ {\bf 45}, S1 (2000).

\bibitem{Noguera:2008}
 S. Noguera and N. N. Scoccola,
 Phys.\ Rev.\  D {\bf 78}, 114002 (2008).

\bibitem{bowman}
P. O. Bowman, U. M. Heller, and A. G. Williams, Phys.
Rev. D {\bf 66}, 014505 (2002);
P. O. Bowman, U. M. Heller, D. B. Leinweber and A. G.
Williams, Nucl. Phys. Proc. Suppl. {\bf 119}, 323 (2003).

\bibitem{Parappilly:2005ei}
  M.~B.~Parappilly, P.~O.~Bowman, U.~M.~Heller, D.~B.~Leinweber,
A.~G.~Williams and J.~B.~Zhang,
  Phys.\ Rev.\ D {\bf 73}, 054504 (2006).

\bibitem {Furui:2006ks}
S.~Furui and H.~Nakajima,
Phys.\ Rev.\ D \textbf{73}, 074503 (2006).

\bibitem{Kondo:2010ts}
  K.~-I.~Kondo,
  Phys.\ Rev.\ D {\bf 82}, 065024 (2010).

\bibitem{Contrera:2010kz}
  G.~A.~Contrera, M.~Orsaria and N.~N.~Scoccola,
  Phys.\ Rev.\ D {\bf 82}, 054026 (2010).

\bibitem{Pagura:2011rt}
  V.~Pagura, D.~Gomez Dumm and N.~N.~Scoccola,
  Phys.\ Lett.\ B {\bf 707}, 76 (2012).

\bibitem{Pagura:2012ku}
  V.~Pagura, D.~Gomez Dumm and N.~N.~Scoccola,
  Phys.\ Rev.\ D {\bf 87}, 014027 (2013).

\bibitem{Hell:2011ic}
  T.~Hell, K.~Kashiwa and W.~Weise,
  Phys.\ Rev.\ D {\bf 83}, 114008 (2011).

\bibitem{Scarpettini:2003fj}
  A.~Scarpettini, D.~Gomez Dumm and N.~N.~Scoccola,
  Phys.\ Rev.\  D {\bf 69}, 114018 (2004).

\bibitem{BB95}
R.~D.~Bowler and M.~C.~Birse,
Nucl. Phys. A {\bf 582}, 655 (1995);
R.~S.~Plant and M.~C.~Birse,
Nucl. Phys. A {\bf 628}, 607 (1998).

\bibitem{L97} H. Leutwyler,
Nucl.\ Phys.\ Proc.\ Suppl.\ {\bf 64}, 223 (1998);
R. Kaiser and H. Leutwyler, in {\it Non-perturbative Methods in
Quantum Field Theory}, edited by A.W. Schreiber, A.G. Williams and
A.W. Thomas (World Scientific, Singapore, 1998),
arXiv:hep-ph/9806336.

\bibitem{F00} T. Feldmann,
Int.\ J.\ Mod.\ Phys.\ A {\bf 15}, 159 (2000).

\bibitem{Beringer:1900zz}
  J.~Beringer {\it et al.}  [Particle Data Group Collaboration],
  Phys.\ Rev.\ D {\bf 86}, 010001 (2012).

\bibitem{Dumitru:2005ng}
  A.~Dumitru, R.~D.~Pisarski and D.~Zschiesche,
  Phys.\ Rev.\  D {\bf 72}, 065008 (2005).

\bibitem{Schaefer:2007pw}
  B.~-J.~Schaefer, J.~M.~Pawlowski and J.~Wambach,
  Phys.\ Rev.\ D {\bf 76 }, 074023 (2007);
  B.~-J.~Schaefer, M.~Wagner and J.~Wambach,
  Phys.\ Rev.\  D {\bf 81 }, 074013 (2010).

\bibitem{GomezDumm:2001fz}
  D.~Gomez Dumm and N.~N.~Scoccola,
  Phys.\ Rev.\ D {\bf 65}, 074021 (2002);
  Phys.\  Rev.\ C {\bf 72}, 014909 (2005).

\bibitem{Borsanyi:2010bp}
  S.~Borsanyi {\it et al.}  [Wuppertal-Budapest Collaboration],
  JHEP {\bf 1009}, 073 (2010).

\bibitem{Bazavov:2010sb}
  A.~Bazavov {\it et al.}  [HotQCD Collaboration],
  J.\ Phys.\ Conf.\ Ser.\  {\bf 230}, 012014 (2010).

\bibitem{Bazavov:2009zn}
  A.~Bazavov {\it et al.},
  Phys.\ Rev.\ D {\bf 80}, 014504 (2009).

\bibitem{Borsanyi:2010cj}
  S.~Borsanyi, G.~Endrodi, Z.~Fodor, A.~Jakovac, S.~D.~Katz, S.~Krieg, C.~Ratti and K.~K.~Szabo,
  JHEP {\bf 1011}, 077 (2010).

\bibitem{Scavenius:2002ru}
  O.~Scavenius, A.~Dumitru and J.~T.~Lenaghan,
  Phys.\ Rev.\ C {\bf 66}, 034903 (2002).

\bibitem{Fukushima:2008wg}
  K.~Fukushima,
  Phys.\ Rev.\ D {\bf 77}, 114028 (2008)
  [Erratum-ibid.\ D {\bf 78}, 039902 (2008)].

\bibitem{Haas:2013qwp}
  L.~M.~Haas, R.~Stiele, J.~Braun, J.~M.~Pawlowski and J.~Schaffner-Bielich,
  Phys.\ Rev.\ D {\bf 87}, 076004 (2013).

\bibitem{Braun:2007bx}
  J.~Braun, H.~Gies and J.~M.~Pawlowski,
  Phys.\ Lett.\ B {\bf 684}, 262 (2010).

\bibitem{Marhauser:2008fz}
  F.~Marhauser and J.~M.~Pawlowski,
  arXiv:0812.1144 [hep-ph].

\bibitem{Herbst:2013ufa}
  T.~K.~Herbst, M.~Mitter, J.~M.~Pawlowski, B.~-J.~Schaefer and R.~Stiele,
  arXiv:1308.3621 [hep-ph].

\bibitem{Costa:2008dp}
  P.~Costa, M.~C.~Ruivo, C.~A.~de Sousa, H.~Hansen and W.~M.~Alberico,
  Phys.\ Rev.\ D {\bf 79}, 116003 (2009).

\bibitem{Fu:2007xc}
  W.~-j.~Fu, Z.~Zhang and Y.~-x.~Liu,
  Phys.\ Rev.\ D {\bf 77}, 014006 (2008).

\bibitem{Sakai:2009dv}
  Y.~Sakai, K.~Kashiwa, H.~Kouno, M.~Matsuzaki and M.~Yahiro,
  Phys.\ Rev.\ D {\bf 79}, 096001 (2009).

\bibitem{Sakai:2010rp}
  Y.~Sakai, T.~Sasaki, H.~Kouno and M.~Yahiro,
  Phys.\ Rev.\ D {\bf 82}, 076003 (2010).

\bibitem{Sasaki:2011wu}
  T.~Sasaki, Y.~Sakai, H.~Kouno and M.~Yahiro,
  Phys.\ Rev.\ D {\bf 84}, 091901 (2011).

\bibitem{Radzhabov:2010dd}
  A.~E.~Radzhabov, D.~Blaschke, M.~Buballa and M.~K.~Volkov,
  Phys.\ Rev.\ D {\bf 83}, 116004 (2011).

\end{thebibliography}
\end{document}